%% file: main.tex
\documentclass[acmsmall]{acmart}
\AtBeginDocument{%
  \providecommand\BibTeX{{%
    \normalfont B\kern-0.5em{\scshape i\kern-0.25em b}\kern-0.8em\TeX}}}
\pdfoutput=1
\input{setup}

\begin{document}

\title{ActRef: Enhancing the Understanding of Python Code Refactoring with Action-Based Analysis}

\author{Siqi Wang}
\affiliation{%
  \institution{Zhejiang University}
  % \streetaddress{30 Shuangqing Rd}
  \city{Hangzhou}
  \state{Zhejiang}
  \country{China}
  }
\email{siqiwang@zju.edu.cn}

\author{Xing Hu}
\authornote{Corresponding Author}
\affiliation{%
  \institution{Zhejiang University}
  % \streetaddress{30 Shuangqing Rd}
  \city{Hangzhou}
  \state{Zhejiang}
  \country{China}}
\email{xinghu@zju.edu.cn}

\author{Xin Xia}
\affiliation{%
  \institution{Zhejiang University}
  % \streetaddress{30 Shuangqing Rd}
  \city{Hangzhou}
  \state{Zhejiang}
  \country{China}
  }
\email{xin.xia@acm.org}

\author{Xinyu Wang}
\affiliation{%
  \institution{Zhejiang University}
  % \streetaddress{30 Shuangqing Rd}
  \city{Hangzhou}
  \state{Zhejiang}
  \country{China}
  }
\email{wangxinyu@zju.edu.cn}

\keywords{Refactoring, Software Maintenance, Python}

\begin{abstract}
  Refactoring, the process of improving the code structure of a software system without altering its behavior, is crucial for managing code evolution in software development. Identifying refactoring actions in source code is essential for understanding software evolution and guiding developers in maintaining and improving the code quality. However, existing tools frequently struggle to detect complex and diverse refactoring types in Python, because most of them are designed for Java-centric environments. This limitation hinders their ability to address Python's dynamic and flexible syntax effectively. This study presents an action-based Refactoring Analysis Framework named \appname, a novel algorithm designed to advance the detection and understanding of Python refactorings through a unique code change action-based analysis of code changes. \appname mining multiple refactoring types (e.g., move, rename, extract, and inline operations) based on diff actions, covering multiple granularity levels including variable, method, class, and module levels.  By focusing on the code change actions, \appname provides a Python-adaptive solution to detect intricate refactoring patterns.
Our evaluation, conducted on 1,914 manually validated refactoring instances from 136 open-source Python projects.
The evaluation results show that \appname achieves high precision (0.80) and recall (0.92), effectively identifying multiple refactoring types. 
Compared with leading baselines, including PyRef, PyRef with MLRefScanner, DeepSeek-R1 and ChatGPT-4, \appname consistently demonstrates superior performance in detecting Python refactorings across various types. While matching PyRef in runtime efficiency, \appname supports a broader spectrum of refactoring types and more refactoring mining levels. \appname shows an effective and scalable approach for mining refactorings in dynamic Python codebases and introduces a new perspective on understanding code.
\end{abstract}

\maketitle

\input{chapter/1-introduction}

\input{chapter/2-Preliminary_Study}
\input{chapter/3-method}

\input{chapter/4-experiment}

\input{chapter/5-RQ}

\input{chapter/6-discussion}
\input{chapter/7-relatedwork}
\input{chapter/8-conclusion}
% \section{DATA AVAILABILITY}
% Both our dataset and source code can be accessed on our website~\cite{Replication-Package} for future research endeavors.

% \newpage

\bibliographystyle{ACM-Reference-Format}
% \IEEEtriggeratref{35}
\bibliography{main}
\end{document}

%% file: setup.tex
\usepackage[ruled,linesnumbered,vlined]{algorithm2e}

\usepackage{caption} 
\usepackage{amsmath}
\usepackage{enumitem}
\usepackage[noend]{algpseudocode}
\usepackage{graphicx}
\usepackage{multicol}
\usepackage{multirow}
\usepackage{makecell}

\usepackage{lineno,hyperref}
\usepackage{booktabs}
\usepackage{calc}
\usepackage{colortbl}
\usepackage{tcolorbox}

\newlength\WIDTHOFBAR
\definecolor{highlightcolor}{RGB}{230, 240, 255}
\definecolor{textcolor}{RGB}{243,104,56} 

\newcommand{\appname}{{\sc ActRef}\xspace}
\newcommand{\appnamebold}{{\sc \textbf{ActRef}}\xspace}

%% file: chapter/1-introduction.tex
\section{Introduction}

Refactoring is a crucial practice in software development that involves the restructuring of existing code without changing its external behavior~\cite{fowler2018refactoring, baqais2020automatic}. 
As software systems grow in scale and complexity, refactoring becomes indispensable for reducing technical debt, improving code readability, reducing complexity, enhancing maintainability, and sustaining long-term software evolution~\cite{maintenance2019,alomar2019impact,alomar2022refactoring,lin2016interactive,iammarino2019self}.
Beyond these immediate benefits, analyzing refactoring practices provides deeper insights into software maintenance, selecting regression tests, good design principles, and patterns of code evolution over time~\cite{wang2018towards, chen2018improving,shen2019intellimerge, DEPALMA2024123602,rachatasumrit2012empirical}.

However, identifying and understanding refactoring operations (e.g., Inline Variable and Extract Class) remains a challenging task. Refactorings are rarely documented systematically by developers~\cite{2019Towards} and are often entangled with other code changes~\cite{silva2016we}. As a result, it becomes difficult to separate refactorings from unrelated changes and to understand their individual impact.
Manual identification of refactorings in large-scale projects is time-consuming and error-prone~\cite{reonciling2012}, further complicating efforts to understand their motivations and impact. To overcome these limitations, several automated tools (e.g., RefactoringMiner~\cite{tsantalis2018accurate,tsantalis2020refactoringminer}, Ref-Finder~\cite{kim2010ref}, and RefDiff~\cite{silva2017refdiff,silva2020refdiff}) are widely used to mine refactoring operations from software projects. These tools have proven effective in statically typed languages such as Java, where type information and rigid syntax help identify structural code changes with high precision.
% These tools are mainly designed for statically typed languages (e.g., Java). They xxx

However, these techniques are not directly applicable to dynamically typed languages like Python~\cite{liu2023automated}, which is the most widely used programming language~\cite{Tiobe2025}.
% In contrast to statically typed languages, Python lacks explicit type annotations, supports runtime type mutation, and allows highly flexible control structures. These features complicate the matching of code elements before and after changes, making it difficult to reliably identify refactoring operations~\cite{vitousek2014design}.
To support Python refactoring detection, two tools have been proposed: Python-Adaptive-RefactoringMiner~\cite{dilhara2022discovering} and PyRef~\cite{atwi2021pyref}, which are both inspired by the Java-based RefactoringMiner. Python-Adaptive-RefactoringMiner translates Python code AST into Java and applies RefactoringMiner’s original rules, but this translation can lose Python-specific semantics and is supported by inferred type information. PyRef avoids translation and applies Java-derived matching mechanisms directly to Python syntax trees. 
Despite their differences, both tools rely heavily on statement-matching, which aligns entire code statements across versions to detect refactorings. Statement-matching is built on the assumption that code structure remains stable across changes and that type information helps disambiguate code elements. While this assumption often holds in statically typed languages, it breaks down in Python, where developers commonly make fine-grained edits within a single statement—for example, modifying a nested expression or changing the structure of a list comprehension~\cite{yang2022mining,ma2022mmf3}. In such cases, statement-matching fails to match fine-grain changed statements, leading to missed or incorrect refactoring detection.

To address these limitations, we propose \appname, a multi-granularity action-based approach for detecting refactoring in Python code. 
Instead of aligning full statements, \appname detects refactorings by analyzing sequences of fine-grained code change actions. Specifically:
\begin{itemize}[leftmargin=*]
    \item To address the fragility of statement-level matching in dynamic languages, \appname builds on and extends GumTree to compute type-agnostic AST-level actions—insertions, deletions, moves, and updates—that tolerate fine-grained edits while preserving semantic intent.
    \item To reduce inaccuracies caused by incorrect AST mappings, \appname incorporates a context-aware post-processing step that refines actions using syntax-aware heuristics and signature consistency. 
    \item To support coarse-grained refactorings' detection, \appname calculates module-level actions that integrate semantic similarity metrics and code slicing techniques across files, allowing it to track code movement and reuse at the module level.
\end{itemize}

Different from existing tools, \appname does not rely on static type information or rigid structure assumptions, which makes it better equipped to handle the flexible features of Python. This makes it highly applicable to dynamic, type-agnostic environments commonly found in Python projects.
Building on this flexibility, \appname provides detailed insights into how code is refactored, offering developers and researchers a structured way to observe, analyze, and understand patterns of code change. By interpreting sequences of code change actions using structured refactoring rules, \appname more closely aligns with a developer's perspective, offering a nuanced approach to uncovering refactoring operations. This action-based method is particularly adept at capturing complex code modification patterns, making it well-suited for dynamic and evolving projects. \appname's ability to track complex code adjustments provides critical insights into the evolution and refinement of Python code, especially valuable in fields like machine learning and data science, where code changes are rapid and complex.

To evaluate the effectiveness of \appname, we compare it with state-of-the-art tools, including PyRef~\cite{atwi2021pyref}, PyRef with MLRefScanner~\cite{noei2025detecting}.
We also compare \appname with two popular Large Language Models (LLMs): DeepSeek-R1~\cite{DeepSeek-R1} and ChatGPT-4~\cite{ChatGPT-4}. We evaluate \appname on a custom-constructed dataset, developed to represent a range of real-world Python refactorings by manually analyzing and addressing limitations found in existing datasets. Our analysis demonstrates that \appname significantly outperforms these baselines in Python refactoring detection across diverse refactoring types, establishing its value for complex, real-world projects. Moreover, \appname's runtime performance matches PyRef's efficiency, while offering a wider range of supported refactorings and enhanced detection granularity. These results underscore \appname's practicality and scalability, presenting it as a robust tool for developers in industry settings. Our paper makes the following contributions:

\begin{itemize}[leftmargin=*]
    \item We propose \appname, an action-based refactoring detection approach specifically designed for Python. \appname proposes a specialized Python refactoring detection strategy, achieving greater performance than state-of-the-art tools.

    \item  To evaluate \appname's effectiveness, we extended a manually analyzed dataset representing a diverse set of real-world Python refactorings, which can be found in our replication package ~\cite{Replication}. This dataset addresses the limitations of existing datasets by capturing complex, real-world refactoring cases, which enhances the relevance and rigor of our evaluation.
\end{itemize}

\noindent\textbf{Paper Organization:} Section \ref{sec:preliminary} describes the motivation and terminologies of our research. Section \ref{sec:method} describes the methodology of our research. Section \ref{sec: setup} describes the evaluation of our search. Section \ref{sec:evaluation} shows the results of our study. Section \ref{sec:discussion} discusses the implications and threats of our results. Section \ref{sec:relatedwork} discusses related work. Section \ref{sec:conclusion} concludes our study and mentions future work.

%% file: chapter/2-Preliminary_Study.tex
\section{Preliminary Study}
\label{sec:preliminary}
In this section, we explain the motivation and terminologies used in our paper.
\subsection{Motivation}

\begin{figure}[h]
    \centering
\includegraphics[width=\linewidth]{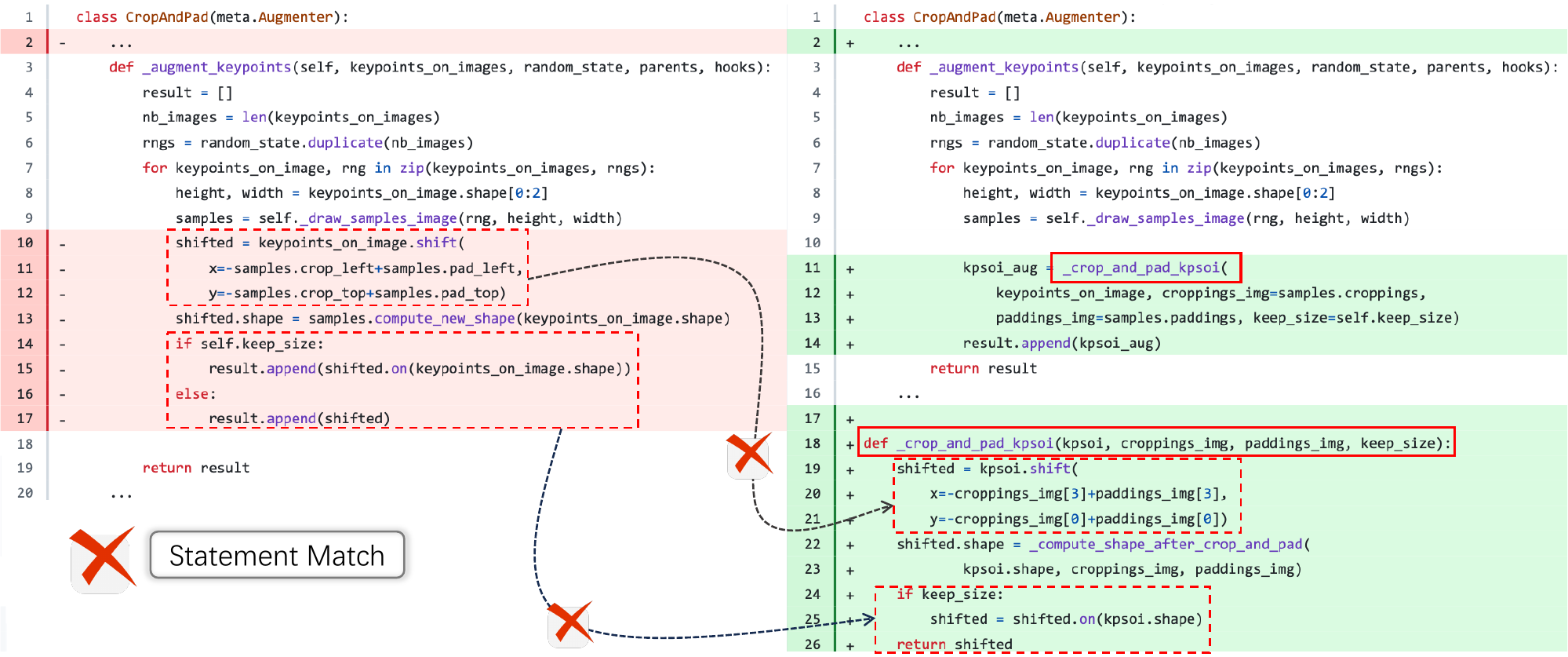}
    \caption{An Extract Method Example from \textbf{imgaug}}
    \label{fig:motivation}
\end{figure}

\begin{figure*}[h]
    \centering
\includegraphics[width=.95\textwidth]{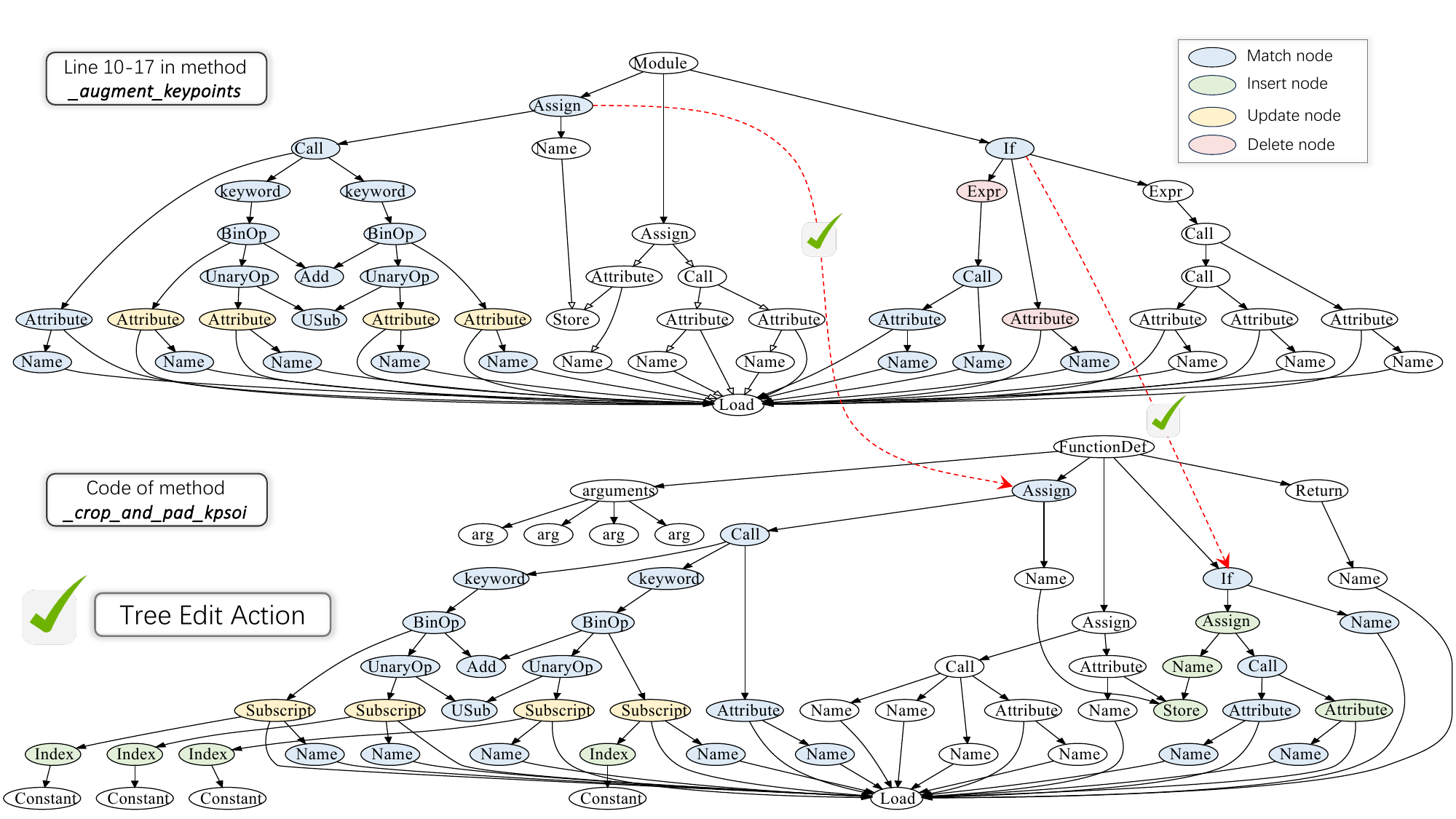}
    \caption{An Extract Method Example AST from \texttt{imgaug}}
    \label{fig:motivation-AST}
\end{figure*}
% \subsection{Motivation Example}

Detecting refactoring in Python code presents unique challenges due to Python’s dynamic typing and its flexible syntax. 
% Unlike statically typed languages like Java, where tools leverage type information to match statements and methods reliably, Python’s lack of type constraints and frequent fine-grained changes can obscure structural similarities during refactoring.
Figure~\ref{fig:motivation} illustrates this challenge with an \textit{Extract Method} example from the \texttt{imgaug}~\cite{Imagaug} project, and Figure~\ref{fig:motivation-AST} provides its corresponding Abstract Syntax Tree (AST). 

In this \textit{Extract Method} example, part of the logic in the \texttt{\_augment\_keypoints} method (lines 10–17) is moved into a newly created method \texttt{\_crop\_and\_pad\_kpsoi}. Existing statement-based tools fail to detect this refactoring due to significant syntax changes and fine-grained variations in the code. The extracted logic introduces a new method with additional parameters, modifies variable assignments, and alters the return mechanism, resulting in statements' structural differences that obscure the relationship between the original and the new code. Moreover, the handling of variables, such as replacing assignments to \texttt{shifted} with direct returns, further highlights the superficial syntactic differences while masking the underlying logical consistency and intent of the developer. 
% \begin{enumerate}
%     \item The extracted logic introduces a new method with additional parameters, modifies variable assignments, and changes the return mechanism.
%     \item Variable handling, such as replacing \texttt{shifted} assignments with direct returns, results in superficial syntactic differences that mask the underlying intent.
% \end{enumerate}

% However, from a developer's perspective, part of the logic in the \texttt{$\_augment\_keypoints$} method (lines 10–17) is extracted into a new method called \texttt{$\_crop\_and\_pad\_kpsoi$}, and the core logic remains unchanged: the sequence of operations—shifting, modifying, and handling \texttt{kpsoi}-is structurally identical in both versions. 

This consistency becomes evident when analyzing the corresponding AST shown in Figure~\ref{fig:motivation-AST}.
Although some AST nodes change during the refactoring, portions of the subtree structure remain consistent, reflecting the underlying logical relationship between the original and the extracted code. From the perspective of AST tree-edit actions, the extraction of logic into a new method can be represented as a combination of ``delete'' subtree actions (removing the logic from the original method) and ``insert'' subtree actions (adding the new method). These actions also capture changes to method signatures, such as parameter additions and return value handling, while preserving the structural flow of the code. This example highlights the limitations of existing tools and emphasizes the importance of analyzing structural consistency through ASTs. Different from statement-matching approaches, which rely heavily on surface-level syntax, the action-based analysis captures the intent behind code changes by leveraging tree-edit operations. 

\subsection{Code Elements}
We extract the needed code elements from AST, including the following elements:

\begin{itemize}[leftmargin=*]
    \item \textbf{Module:} A module is defined as a Python file, containing classes and methods, and directly affiliated definitions and statements (i.e., those not defined within a method).
    \item \textbf{Class:} A class is a blueprint for creating objects, encapsulating methods, statements, and variables. It supports object-oriented features such as inheritance, encapsulation, and polymorphism. Classes are defined with the `\texttt{class}' keyword in Python.
    \item \textbf{Method:} A method is a reusable block of code associated with a class or module, typically performing a specific operation. It consists of a declaration, an optional list of parameters, and a suit of executable statements. Methods are defined with the `\texttt{def}' keyword in Python.
    % \item \textbf{Suit:} A suit is a block of indented statements that execute together, forming the body of a construct such as a method, loop, conditional, or class definition. For example, in a method, the suit represents its core executable logic.
    \item \textbf{Statement:} A statement is the smallest executable unit in Python, representing a single action or command. It is terminated by a line break or a semicolon and can be categorized into simple statements (e.g., assignments and expressions) or compound statements (e.g., conditionals, loops, and function definitions).
    % \item \textbf{Statement:} A statement is the smallest execution unit in Python and is terminated by a line break or a semicolon. It includes many types (e.g., conditional statement and expression) 
    \item \textbf{Variable:} A variable is an identifier used to store data in memory. It is the smallest nameable storage element in a Python program, capable of referencing various data types such as integers, strings, or objects.
\end{itemize}

\subsection{Action}

In code refactoring detection, \textbf{action} represents the atomic changes that occur during the modification of code elements. These actions serve as the foundation for understanding how code evolves during refactoring, capturing both structural and semantic intentions behind each change. Rather than relying solely on statement-matching or token diffing, our method computes multi-level actions that integrate both fine-grained and coarse-grained perspectives on code changes.
To capture these actions, we design a multi-level action modeling strategy based on the granularity of code transformation:
\begin{itemize}[leftmargin=*]
    \item \textbf{AST-level actions:}
% For intra-file and fine-grained code changes, we utilize the self-adaptive heuristic matcher of GumTree~\cite{falleri2014fine,falleri2024fine}, a state-of-the-art AST differencing algorithm based on self-adaptive subtree matching. It identifies minimal edit operations required to transform the AST of the original file into the AST of the modified file. These actions follow the representation adopted in prior work~\cite{chawathe1996change,hashimoto2008diff,falleri2014fine}, with the following atomic types:

For fine-grained changes, we analyze differences between the abstract syntax trees (ASTs) of the original and modified files. These actions are enriched with contextual and semantic information, including the signatures of methods or classes and their surrounding code structure. The output consists of atomic operations such as:
\begin{itemize}[leftmargin=*]
\item \textbf{delete}$(d)$: delete code element $d$.
\item \textbf{insert}$(i, n)$: insert code element $i$ at position $n$.
\item \textbf{move}$(m, o, n)$: move code element $m$ from position $o$ to $n$.
\item \textbf{update}$(o, n)$: change the code value from $o$ to $n$.
\end{itemize}

We build our AST-level action calculator on top of GumTree~\cite{falleri2014fine,falleri2024fine}, a widely used AST differencing engine, and enhance it with semantic-aware extensions that consider code signatures, usage context, and cross-file coordination.
This action set provides the foundation for detecting intra-file and cross-file fine-grained refactorings (class/method/variable level).
% These AST-level actions are computed using a hybrid approach that combines a structural differencing algorithm with custom logic that examines code semantics and usage contexts. While we use GumTree~\cite{falleri2014fine,falleri2024fine} as the base differencing engine to identify candidate edits, we extend and refine its output with domain-specific rules, semantic context integration, and multi-file coordination.

% \item\textbf{Slice-level actions:}
\item\textbf{Module-level actions:}
% We adopt a slicing-based strategy to infer coarse-grained actions at the module level, where structural matching via AST may be insufficient, such as in file-level additions, deletions, or large-scale refactorings. Each file is first divided into top-level code slices, separated by double newlines, which often correspond to function definitions, class declarations, or logical blocks. We then compare these code slices across the before and after versions using text similarity metrics. If a code slice in the new file exhibits high similarity to a deleted code slice in another file, we treat this as a potential module-level \textbf{move} action. These inferred actions of large code fragments are essential for detecting higher-level operations such as \textit{Extract Module} and \textit{Inline Module}.
For coarse-grained code transformations (e.g., moving or extracting large code blocks across modules), we introduce \textit{module-level actions}. We build our module-level action calculator combining program slicing and similarity to overcome the limitations of AST matching in scenarios involving entire file additions, deletions, or large-scale relocations. We define the following atomic operations at the coarse-grained refactoring (module level):
\begin{itemize}[leftmargin=*]
        \item \textbf{insert}$(s, f)$: code slice $s$ is newly added to file $f$.
        \item \textbf{delete}$(s, f)$: code slice $s$ is deleted from file $f$.
        \item \textbf{move}$(s, f_o, f_n)$: code slice $s$ is moved from $f_o$ to $f_n$.
\end{itemize}
% These actions enable us to detect the coarse-grained refactoring (module level).
\end{itemize}

By analyzing both fine-grained AST-level and coarse-grained module-level actions, our multi-level action analysis framework enables a broader and more accurate coverage of refactoring detection. This design not only improves fidelity to developer behavior but also enhances the robustness of refactoring detection in dynamic languages such as Python.

 % By analyzing these atomic actions, our method seeks to understand code refactoring in a way that aligns more closely with developers' workflows, as they often refactor code through non-linear, cross-file updates that may not match precisely on a node-by-node or statement-by-statement level. The action-based method is more flexible and is capable of capturing cross-file and cross-level refactoring operations.

%% file: chapter/3-method.tex
\section{Methodology}
\label{sec:method}

In this section, we introduce an action-based approach to detect code refactoring operations. 
Unlike existing methods that rely on statement-level matching, \appname focuses on action, which can indicate the developer’s intent, treating changes as purposeful operations such as ``\textit{moving code to extract a new method}’’.
By capturing refactorings from this perspective, \appname offers a more intuitive and semantically meaningful view of code evolution.

%It can be reasonably assumed that 
% Software developers typically consider the actions they perform, such as moving methods, extracting classes, or renaming variables, as opposed to purely structural transformations. 
% By concentrating on these actions, \appname is designed to correspond more closely with the cognitive processes of software developers. 

 % This involves the analysis of the actions generated during the code modification process, including insertion, deletion, and movement, to identify the refactoring patterns present in the code. The prevailing approach relies on the precise matching of code nodes~\cite{tsantalis2020refactoringminer}, which is inadequate for capturing the complex and nonlinear modifications that developers typically make during the refactoring process.
% \subsection{Action-Based Detection}
\begin{figure}
    \centering
    \includegraphics[width=0.8\linewidth]{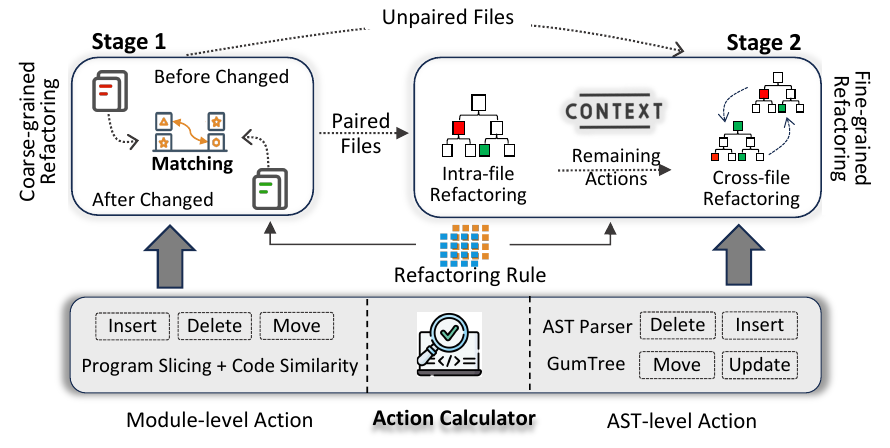}
    \caption{Methodology Overview}
    \label{fig:overview}
\end{figure}

 \subsection{Overall Framework}
Figure~\ref{fig:overview} shows the overview framework of our approach \appname. By analysing  Sets of files before and after a commit, we detect both coarse-grained refactorings (module level) and fine-grained refactorings (class/method/variable level) in this commit.
It involves decomposing code modifications into a series of actions, using staged refactoring mining to identify refactoring. We employ a model-level action calculator and an AST-level action calculator to get the code change actions. By further matching these actions with predefined refactoring rules, it is possible to accurately identify 15 types of refactoring operations (e.g., Extract Variable, Rename Module, Move Method). This section provides a comprehensive overview of the algorithmic design, detection rules, and the rationale behind \appname.

\begin{algorithm}
\footnotesize

\SetAlgoNoEnd 
\caption{Overall Algorithm of Refactoring Detection}\label{algo:overall}

\KwIn{$\mathcal{F}^-, \mathcal{F}^+$: Sets of files before and after a commit}
\KwOut{$\mathcal{R}$: Detected Refactorings}

\BlankLine
    $\mathcal{R} \gets \emptyset$ \tcp*{Initialize refactoring set  }
    $\mathcal{I}_{\text{unpaired}} \gets \emptyset, \mathcal{D}_{\text{unpaired}} \gets \emptyset$ \tcp*{Unpaired insert/delete actions for cross-file matching}
\BlankLine
\tcp{Step 1: Coarse-grained refactoring detection}
    $\mathcal{P}, \mathcal{D}_{\text{unpaired}}, \mathcal{I}_{\text{unpaired}}, \mathcal{R}_{\text{module}} \gets \text{ModuleLevelRefactoringDetection}(\mathcal{F}^-, \mathcal{F}^+)$
\BlankLine
\tcp{Step 2: Fine-grained refactoring detection}
\For{$(f_i^-, f_i^+) \in \mathcal{P}$}{
    % $Action_i \gets \text{ActionCalculation}(f_i^-, f_i^+)$\;
    $\mathcal{R}_{\text{intra}}, Action_i^{\text{rem}} \gets \text{IntraFileRefactoringDetection}(f_i^-, f_i^+)$\;
}
$Action_{\text{remaining}} \gets Action_1^{rem} \cup \dots \cup  Action_n^{rem}  $
\BlankLine
% \tcp{Step 3: Cross-file refactoring detection}
$\mathcal{R}_{\text{cross}} \gets \text{CrossFileRefactoringDetection}(\mathcal{I}_{\text{unpaired}}, \mathcal{D}_{\text{unpaired}}, Action_{\text{remaining}})$\;

$\mathcal{R} \gets \mathcal{R}_{\text{intra}} \cup \mathcal{R}_{\text{cross}} \cup \mathcal{R}_{\text{module}}$\;

\Return{$\mathcal{R}$}\;
\end{algorithm}

% \begin{algorithm}
% \footnotesize
% % \SetAlgoNoEnd 
% \caption{Refactoring Detection Overall Algorithm}\label{algo:overall}
% \KwIn{$F_{\text{before}}, F_{\text{after}}$: Two sets of files}
% \KwOut{$R$: Detected Refactorings}
% \BlankLine
% % \SetKwProg{Fn}{Function}{:}{}
% % \Fn{AnalyseParameterTransferType($methodCallPath$)}{
%     $R \gets \emptyset$ \tcp*{Initialize refactoring set  }
%     $I_{\text{unpaired}} \gets \emptyset, D_{\text{unpaired}} \gets \emptyset$ \tcp*{Unpaired actions for cross-file matching}
% \BlankLine
% \tcp{Step 1: Intra-file calculation for each pair of files}
%     $P, D_{\text{unpaired}}, I_{\text{unpaired}}, R_{\text{module}} \gets \text{CompareFiles}(F_{\text{before}}, F_{\text{after}})$
%     \For{$(f_{\text{before}}, f_{\text{after}}) \in P$}{
%     $Actions \gets \text{ActionCalculation}(f_{\text{before}}, f_{\text{after}})$\;
%     $R_{\text{intra}},  Actions_{\text{remaining}} \gets \text{IntrafileRefactoringDetection}(Actions))$\;
%     % $I_{\text{unpaired}} \gets I_{\text{unpaired}} \cup I$\;
%     % $D_{\text{unpaired}} \gets D_{\text{unpaired}} \cup D$\;
%     }
% \BlankLine
%     \tcp{Step 2: Cross-file refactoring detection}
%     $R_{\text{cross}} \gets \text{CrossfileRefactoringDetection}(I_{\text{unpaired}}, D_{\text{unpaired}}, Actions_{\text{remaining}})$\;
%     $R \gets R_{\text{intra}} \cup R_{\text{cross}}\cup R_{\text{module}}$\;
%     \Return{$R$}\;
% % }
% \end{algorithm}

As shown in algorithm~\ref{algo:overall}, the input of \appname is two sets of files, designated as $\mathcal{F}^-$ and $\mathcal{F}^+$, which represent the state before and after the code is modified. 
We first detect the coarse-grained refactoring, the function \texttt{ModuleLevelRefactoringDetection} is responsible for pairing up files and detecting module-level refactorings from $\mathcal{F}^-$ and $\mathcal{F}^+$ based on their names, paths, and code similarity. It returns four sets:
(1) $\mathcal{P}$: The set of paired files, (2)
$\mathcal{D}_{\text{unpaired}}$: Deletions that exist only in $\mathcal{F}^-$ (deleted files), (3)
$\mathcal{I}_{\text{unpaired}}$: Insertions that exist only in $\mathcal{F}^+$ (newly added files) and (4) $\mathcal{R}_{\text{module}}$: module-level refactorings. Then, we detect the fine-grained refactorings. We generate the intra-file code change actions using AST-level action calculator and process these actions to mining refactoring operations. Subsequently, the unmatched insert and delete actions, which have not been calculated in the intra-file detection step, are passed to the cross-file detection step. Finally, we combine the remaining actions and unpaired files, and generate the cross-file code change actions using AST-level action calculator to detect cross-file refactorings.

% \begin{center}
% \begin{minipage}{.8\linewidth}

% \end{minipage}
% \end{center}

\subsection{Coarse-grained Refactoring Detection}
\begin{algorithm}
\footnotesize
\SetAlgoNoEnd 
\caption{Module-level Refactoring Detection}
\label{algo:modulelevel}

\KwIn{$\mathcal{F}^-, \mathcal{F}^+$: Sets of files before and after a commit}
\KwOut{$\mathcal{R}_{\text{module}}$: Detected module-level refactorings}

\BlankLine
$\mathcal{R}_{\text{module}} \gets \emptyset$ \\
$\mathcal{P} \gets \emptyset$, $\mathcal{D}_{\text{unpaired}} \gets \emptyset$, $\mathcal{I}_{\text{unpaired}} \gets \emptyset$ \\
\BlankLine

\tcp{Step 1: Match files using name, path, content similarity}
\For{$f_b \in \mathcal{F}^-, f_a \in \mathcal{F}^+$} {
    \If{$\text{is\_same\_name}(f_b, f_a)$} {
        $\mathcal{P} \gets \mathcal{P} \cup (f_b, f_a)$
    }
    \ElseIf{$(f_a, f_b)$ can match rule in Table1} {
        $\mathcal{R}_{\text{module}} \gets \mathcal{R}_{\text{module}} \cup $ corresponding refactoring\;
        $\mathcal{P} \gets \mathcal{P} \cup (f_b, f_a)$
    }

    \Else{
    $\mathcal{D}_{\text{unpaired}} \gets f_b$\ , $\mathcal{I}_{\text{unpaired}} \gets f_a$\;
    }
}

\BlankLine
\tcp{Step 2: Detect Extract/Inline Module}
\For{$f_a \in \mathcal{I}_{\text{unpaired}}$} {
    \If{$f_a$ can match rule in Table1 }{
        $\mathcal{R}_{\text{module}} \gets \mathcal{R}_{\text{module}} \cup \text{ExtractModule}(f_a)$\;
        Remove $f_a$ from $\mathcal{I}_{\text{unpaired}}$
    }
}
\For{$f_b \in \mathcal{D}_{\text{unpaired}}$} {
    \If{$f_b$ can match rule in Table1 }{
        $\mathcal{R}_{\text{module}} \gets \mathcal{R}_{\text{module}} \cup \text{InlineModule}(f_b)$\;
        Remove $f_b$ from $\mathcal{D}_{\text{unpaired}}$
    }
}

\Return{$\mathcal{R}_{\text{module}}, \mathcal{D}, \mathcal{D}_{\text{unpaired}}, \mathcal{I}_{\text{unpaired}}$}

\end{algorithm}
Algorithm~\ref{algo:modulelevel} aims to detect coarse-grained refactorings. The input to this procedure is the complete set of files before ($\mathcal{F}^-$) and after ($\mathcal{F}^+$) a commit. The algorithm proceeds in two main stages:
The first step iterate over all possible pairs $(f_b, f_a)$, where $f_b \in \mathcal{F}^-$ and $f_a \in \mathcal{F}^+$. If the two files share the same name and path, they are directly paired and added to the matched file set $\mathcal{P}$. Otherwise, we check whether the pair $(f_b, f_a)$ satisfies any module-level transformation rules defined in Table~\ref{tab:actref-rules} (e.g., Rename Module, Move Module). If a rule matches, the corresponding refactoring is recorded and the pair is also marked as matched. Files that do not satisfy these rules are considered unmatched and added to the sets $\mathcal{D}_{\text{unpaired}}$  and $\mathcal{I}_{\text{unpaired}}$.
The second step capture extract and inline module patterns, we analyze the unmatched files collected in the previous step. For each unpaired file, we check whether it satisfies Extract Module or Inline Module rule. If such a match exists, we add the corresponding refactoring operation to $\mathcal{R}_{\text{module}}$ and remove this file from unpaird file set.

% The output of this algorithm includes the set of detected module-level refactorings $\mathcal{R}_{\text{module}}$, along with the remaining unmatched files, which may be further analyzed in cross-file detection or reported as potential file additions or deletions.

% \begin{center}

\begin{table}[htbp] \small

\caption{Detection Rules for Refactoring Types}
\label{tab:actref-rules}
\resizebox{.95\linewidth}{!}{
\begin{tabular}{m{2.5cm}|m{11.7cm}}

\toprule

\multicolumn{1}{c}{\textbf{Refactoring}} & \multicolumn{1}{c}{\textbf{Rule}} \\ \toprule

\centering\arraybackslash Extract Module & \setlength\leftskip{1.5em}
$\exists f \in \mathcal{F}^+ , \mathsf{slices}(f) = {s_1, \dots, s_n} ,  \forall s_i \mid is\_move\_silce(s_i) $
\\ \midrule

% \centering\arraybackslash {Extract Module} & 
% \begin{enumerate}[leftmargin=*,after=\vspace{-\baselineskip}]
%     \item  $\exists f \in \text{delete\_actions}, i \in \text{insert\_actions}, c \in \text{insert\_actions}  
%     \mid \text{is\_similar}(\text{body}(i), d) \wedge \text{is\_class(c)}$
% \item $\exists m \in \text{move\_actions}, i \in \text{insert\_actions}, c \in \text{insert\_actions} \mid \text{is\_class}(\text{c}) \wedge \text{same\_name}(i, c) \wedge \text{is\_body(m, c)}$ 

% \end{enumerate} 
% \\
% \midrule

\centering\arraybackslash {Extract Class} & 
\begin{enumerate}[leftmargin=*,after=\vspace{-\baselineskip}]
    \item  $\exists d \in \text{delete\_actions}, i \in \text{insert\_actions}, c \in \text{insert\_actions}  
    \mid \text{is\_similar}(\text{body}(i), d) \wedge \text{is\_class(c)}$
\item $\exists m \in \text{move\_actions}, i \in \text{insert\_actions}, c \in \text{insert\_actions} \mid \text{is\_class}(\text{c}) \wedge \text{same\_name}(i, c) \wedge \text{is\_body(m, c)}$ 

\end{enumerate} 
\\
\midrule

\centering\arraybackslash Extract Method & 
\begin{enumerate}[leftmargin=*,after=\vspace{-\baselineskip}]
\item $\exists \, d \in \text{delete\_actions}, \, i \in \text{insert\_actions}, \, c \in \text{insert\_actions} \mid \text{is\_similar}(\text{body}(i), d) \wedge \text{is\_method}(i)  $
\item $\exists \, m \in \text{move\_actions}, \, i \in \text{insert\_actions}, \, c \in \text{insert\_actions} \mid   \text{is\_method}(i)  \wedge \text{same\_name}(i, c)$ 
\end{enumerate}
  \\
\midrule

\centering\arraybackslash Extract Variable & 
\begin{enumerate}[leftmargin=*,after=\vspace{-\baselineskip}]
\item $\exists \, d \in \text{delete\_actions}, \, i \in \text{insert\_actions}, \, c \in \text{insert\_actions}
 \mid \text{is\_similar}(\text{RHS}(i), d) \wedge \text{is\_variable}(i)$
\item$ \exists\, m \in \text{move\_actions},\, i \in \text{insert\_actions}, \, c \in \text{insert\_actions}  \mid
\text{is\_expression}(m)$
\end{enumerate}
  \\
\midrule

\centering\arraybackslash Move Module &    \setlength\leftskip{1.5em} {$\exists f_b \in \mathcal{F}^-, f_a \in \mathcal{F}^+ \mid (f_b, f_a) \notin \mathcal{P} \wedge  \text{is\_similar}(f_a, f_b) \wedge \text{different\_path}(f_b, f_a)$} \\\midrule

\centering\arraybackslash Move Class & 
\begin{enumerate}[leftmargin=*,after=\vspace{-\baselineskip}]
\item $\exists \, m \in \text{move\_actions}  \mid  \text{is\_class}(m) $
\item$\exists d \in \text{delete\_actions}, i \in \text{insert\_actions}  \mid  \text{is\_class}(d) \wedge \text{is\_class}(i) \wedge \text{is\_similar}(d,i)$
\end{enumerate}
  \\
\midrule

\centering\arraybackslash Move Method & 
\begin{enumerate}[leftmargin=*,after=\vspace{-\baselineskip}]
\item $\exists \, m \in \text{move\_actions}  \mid  \text{is\_method}(m) $ 
\item$\exists d \in \text{delete\_actions}, i \in \text{insert\_actions} \mid  \text{is\_method}(d) \wedge \text{is\_method}(i) \wedge \text{is\_similar}(d,i)$
\end{enumerate}
  \\
\midrule

\centering\arraybackslash Inline Module & 
\setlength\leftskip{1.5em}$\exists f \in \mathcal{F}^- , \mathsf{slices}(f) = {s_1, \dots, s_n} , \forall s_i \mid is\_move\_silce(s_i) $
\\ \midrule

\centering\arraybackslash Inline Class & 
\begin{enumerate}[leftmargin=*,after=\vspace{-\baselineskip}]
\item $\exists \, d \in \text{delete\_actions}, \, c \in \text{delete\_actions}, \, i \in \text{insert\_actions} \mid  \text{is\_similar}(\text{body}(d), i) \wedge \text{is\_class}(d)  $
\item $\exists \, m \in \text{move\_actions}, \, d \in \text{delete\_actions}, \, c \in \text{delete\_actions} \mid
  \text{is\_class}(d) \wedge \text{same\_name}(d, c)$
\end{enumerate}
  \\
\midrule

\centering\arraybackslash Inline Method & 
\begin{enumerate}[leftmargin=*,after=\vspace{-\baselineskip}]
\item $\exists \, d \in \text{delete\_actions}, \, c \in \text{delete\_actions}, \, i \in \text{insert\_actions}  \mid
 \text{is\_similar}(\text{body}(d), i) \wedge \text{is\_method}(d) $
\item $\exists \, m \in \text{move\_actions}, \, d \in \text{delete\_actions}, \, c \in \text{delete\_actions} \mid
  \text{is\_method}(d)  \wedge \text{same\_name}(d, c)$
\end{enumerate}
  \\
\midrule

\centering\arraybackslash Inline Variable & 
\begin{enumerate}[leftmargin=*,after=\vspace{-\baselineskip}]
\item $\exists \, d \in \text{delete\_actions}, \, i \in \text{insert\_actions}, \, c \in \text{delete\_actions}
  \mid \text{is\_similar}(\text{RHS}(d), i) \wedge \text{is\_variable}(i) $
\item $\exists\, m \in \text{move\_actions},\, d \in \text{delete\_actions}, \, c \in \text{delete\_actions}  \mid
\text{is\_expression}(m)$
\end{enumerate}
  \\
\midrule

Rename Module &     \setlength\leftskip{1.5em}{$\exists f_b \in \mathcal{F}^-, f_a \in \mathcal{F}^+ \mid (f_b, f_a) \notin \mathcal{P} \wedge  \text{is\_similar}(f_a, f_b) \wedge \text{different\_name}(f_b, f_a)$} \\\midrule

Rename Class & \setlength\leftskip{1.5em}\colorbox{highlightcolor}{\textcolor{textcolor}{$\exists u \in \text{update\_actions} \wedge \text{is\_class\_name}(u)$}} \\\midrule

Rename Method & \setlength\leftskip{1.5em}\colorbox{highlightcolor}{\textcolor{textcolor}{$\exists u \in \text{update\_actions} \wedge \text{is\_method\_name}(u)$}} \\\midrule

Rename Variable &\setlength\leftskip{1.5em}\colorbox{highlightcolor}{\textcolor{textcolor}{$\exists u \in \text{update\_actions} \wedge \text{is\_expression\_LHS}(u)$}} \\

\bottomrule
\end{tabular}}
\begin{quote}
    \footnotesize{* LHS: the left hand side of an expression statement. }
\end{quote}
\end{table}

\subsection{Fine-grained Refactoring Detection}
Algorithm~\ref{algo:intrafile} and Algorithm~\ref {algo:crossfile} aim to detect fine-grained refactorings.
\subsubsection{Intra-file Refactoring Detection}
\begin{algorithm}
\footnotesize

\SetAlgoNoEnd 
\caption{Intra-file Refactoring Detection}\label{algo:intrafile}

\KwIn{$f_i^-, f_i^+$ \tcp*{Matched file pair}}
\KwOut{$\mathcal{R}_{\text{intra}}$: Detected refactorings}

\BlankLine
% \SetKwProg{Fn}{Function}{:}{}
% \Fn{AnalyseParameterTransferType($methodCallPath$)}{
    % $R \gets \emptyset$ \tcp*{Initialize refactoring set  }
    % $Actions \gets \emptyset$ \tcp*{Initialize actions set  }
    % $I_{\text{unpaired}} \gets \emptyset, D_{\text{unpaired}} \gets \emptyset$ \tcp*{Unpaired actions for cross-file matching}
    % $R \gets \emptyset$  , $Actions \gets \emptyset$ \tcp*{Initialize refactoring set and actions set  }
    
\BlankLine
$\mathcal{R}_{\text{intra}} \gets \emptyset$, $Action_i \gets \emptyset$, $ Acions_i^{\text{rem}}$ \tcp*{Initialize refactoring and action sets}

\BlankLine
\tcp{Step 1: Action Generation}
$AST_i^- \gets \texttt{GenerateAST}(f_i^-)$ \\
$AST_i^+ \gets \texttt{GenerateAST}(f_i^+)$ \\

$Action_i \gets \texttt{ActionCalculation}(AST_i^-, AST_i^+)$

\BlankLine
    \tcp{Step 2: Intra-file Refactoring Detection:}
    \For{$\alpha \in Action_i$}{

    \If{$\alpha$ is \textit{move} or \textit{update}}{
        
        \If{$\alpha$ can match rule in Table 1}{
            $\mathcal{R}_{\text{intra}} \gets \mathcal{R}_{\text{intra}} \cup$ corresponding refactoring\;
            Remove $\alpha$ from $Action_i$\;
        }
    }
    % \If{$\alpha$ is an \textit{insert} or \textit{delete}}{
    % \If{rule for $\alpha$ found in refactoring table}{
    %         $R_{\text{intra}} \leftarrow R_{\text{intra}} \cup \text{corresponding refactoring}$\;
    %         Remove corresponding actions from $Acions$\;
    %     }
    % }
    }
$Actions_i^\text{rem} \gets Actions_i $\\    

\Return{$R_{\text{intra}} , 
 Actions_i^\text{rem}$}

\end{algorithm}

% \end{center}

Algorithm~\ref{algo:intrafile} aims to detect intra-file refactorings based on fine-grained edit actions between two versions of the same file. The input to the algorithm is a pair of files, designated as $f_{\text{before}}$ and $f_{\text{after}}$, representing the files before and after the change. The initial step (lines 2-3) generates an abstract syntax tree of the input file and entails a comparison of the abstract syntax trees of the preceding and subsequent versions to calculate specific change actions. These actions encompass the \textit{insert}, \textit{delete}, \textit{move}, and \textit{update} of code. 
After obtaining the set of edit actions, we match them against a predefined set of refactoring rules as shown in Table~\ref{tab:actref-rules}. Each rule defines a structural pattern of actions that together constitute a recognizable refactoring operation.
Once a rule is successfully matched, we add the corresponding refactoring operation $r$ to the set $R_{\text{intra}}$, and remove all actions involved in the match from the working action set to avoid duplication in subsequent rule matches.
% These actions serve as the foundation for subsequent detection of refactoring types. Subsequently, step 2 (lines 6-12) provides a detailed examination and calculation of the actions identified in step 1, to identify the specific refactoring operations. This step encompasses the application of specific processing rules to a range of actions. To illustrate, upon the detection of a move or update action (line 5), the algorithm proceeds to ascertain the source and target nodes (lines 6). In the event that the move pertains to a class, method, or variable (line 7), the action is identified as the corresponding refactoring type. Consequently, the corresponding move, insert, and delete actions are removed from the action set (line 9). Similarly, in the case of insert or delete actions (line 10), if the corresponding rules are identified in the refactoring rule table~\ref{tab:actref-rules}, these actions are marked as the corresponding refactoring types  and the remaining action set is updated (lines 12-13).

% \begin{center}
%     \begin{minipage}{.8\linewidth}
        
% \algtext*{End}

\subsubsection{Cross-file Refactoring Detection}

\begin{algorithm}
\footnotesize
\SetAlgoNoEnd 
\caption{Cross-file Refactoring Detection}\label{algo:crossfile}

\KwIn{$\mathcal{D_{\text{unpaired}}}, \mathcal{I_{\text{unpaired}}}, Acions_{\text{remaining}}$}
\KwOut{$\mathcal{R_{\text{cross}}}$}

\BlankLine
% \SetKwProg{Fn}{Function}{:}{}
% \Fn{AnalyseParameterTransferType($methodCallPath$)}{

$\mathcal{R}_{\text{cross}} \gets \emptyset$ \tcp*{Initialize set for cross-file detection}
\For{$f_{\text{del}} \in \mathcal{D_{\text{unpaired}}}$}{
    Convert $f_{\text{del}}$ to a delete action and add it to $Actions_{\text{remaining}}$ \;
}
\ForEach{$f_{\text{add}} \in \mathcal{I_{\text{unpaired}}}$}{
    Convert $f_{\text{add}}$ to an add action and add it to $Actions_{\text{remaining}}$ \;
}
\For{pair of delete action $\alpha_{\text{del}} \in Acions_{\text{remaining}}$ and add action $\alpha_{\text{add}} \in Actions_{\text{remaining}}$}{
    Extract code snippets $C_{\text{del}}$ from $\alpha_{\text{del}}$ and $C_{\text{add}}$ from $\alpha_{\text{add}}$ \;
    Compute actions between $C_{\text{del}}$ and $C_{\text{add}}$ using GumTree \;
    \If{actions satisfy refactoring rules}{
        Add the detected refactoring to $R_{\text{cross}}$ \;
        Remove matched actions from $Actions_{\text{remaining}}$ \;
    }
}
\Return{$R_{\text{cross}}$}
\end{algorithm}
%     \end{minipage}
% \end{center}

Algorithm ~\ref{algo:crossfile} outlines the algorithm for cross-file refactoring detection, which handles code changes across file boundaries. The input of cross-file refactoring detection includes unpaired files and remaining actions. For the unpaired deleted file set $D\_unpaired$ (lines 2-3), each file is transformed into a deletion action and incorporated into the set of remaining actions. Similarly, the unpaired add file set $I\_unpaired$ (lines 4-5) is treated analogously, with each file converted to an add action and processed accordingly.
The central processing step (lines 6-11) entails an iterative process involving the remaining pairs of delete and add actions. For each pair of actions (lines 7-8), the algorithm extracts the pertinent code segments and employs the AST-level action calculator to get the precise actions between the two code segments. If the actions align with the predefined refactoring rules (line 9) shown in table~\ref{tab:actref-rules}, they are identified as a valid cross-file refactoring, and this refactoring is incorporated into $R\_cross$ (line 10). Upon completion of the aforementioned steps, the matched actions are removed from the action set (line 11) to avoid any potential for duplicate processing.

%% file: chapter/4-experiment.tex
\section{Evaluation}
\label{sec: setup}
\subsection{Baselines}
To evaluate effectiveness of \appname in detecting Python refactorings, we compare it with the following baseline tools: \textbf{PyRef}~\cite{atwi2021pyref}, \textbf{PyRef} with \textbf{MLRefScanner}~\cite{noei2025detecting}, \textbf{DeepSeek-R1}~\cite{DeepSeek-R1}, and \textbf{ChatGPT-4}~\cite{ChatGPT-4}. PyRef is the state-of-the-art rule-based refactoring detection tool, specifically from RefactoringMiner2~\cite{tsantalis2020refactoringminer} adapted for Python, which supports several 
key refactoring types (e.g., Move Method, Inline Method, and Rename Class).
MLRefScanner is a classifier to detect python refactoring commits, which can ensemble with PyRef. 
On the other hand, DeepSeek-R1 and ChatGPT-4 are the LLMs pre-trained on diverse codebases, offering the flexibility of pattern recognition informed by general knowledge of refactoring.
Adaptive-Python-Refactoringminer~\cite{dilhara2022discovering} was considered, we excluded it due to replication challenges.
% Although we initially considered using Python-Adaptive-Refactoring Miner~\cite{dilhara2022discovering} and were able to successfully execute its replication package with assistance from the original authors, we ultimately excluded it from our evaluation due to significant reliability issues. Specifically, the tool failed to detect any refactoring operations, even when applied to the original dataset provided by the authors. Furthermore, this limitation is not unique to our experience, similar issues have been reported by other users on the tool’s official replication package repository\footnote{https://github.com/mlcodepatterns/PythonTypeInformation/issues/2}, suggesting a broader reproducibility problem. 

\subsubsection{PyRef}

PyRef~\cite{atwi2021pyref} is the state-of-the-art tool that automatically detects method-level refactoring operations in Python projects. The tool was inspired by RefactoringMiner~\cite{tsantalis2020refactoringminer}, using the same AST-based statement matching algorithm that determines refactoring candidates without requiring user-defined thresholds. Although RefactoringMiner2 was originally developed for Java, PyRef incorporates Python's syntax and semantics to support accurate refactoring detection in Python projects. However, PyRef only supports several key refactoring types in method-level refactoring and extended Rename Class recently.

\subsubsection{MLRefScanner}

MLRefScanner~\cite{noei2025detecting} is a prototype tool that applies ML techniques to detect refactoring commits in ML Python projects. MLRefScanner extracts textual, process, and code features to train classifiers for detecting commits. However, MLRefScanner relies on high-quality commit messages and is unable to detect specific types of refactorings. 
In the original setting, MLRefScanner and PyRef are combined by taking the union of refactoring-labeled commits, enabling coarse-grained (binary) detection. However, this approach cannot provide specific refactoring types and is thus not suitable for fine-grained evaluation.
To make this baseline comparable to our method, we take the intersection of commits detected by both MLRefScanner and PyRef. We then use PyRef’s output to label the specific refactoring operations. This ensures that the baseline only reports refactorings with higher confidence (both tools agree) and includes type-level details, enabling a fair comparison.

\subsubsection{DeepSeek-R1 and ChatGPT-4}
DeepSeek-R1 is an open-source LLM trained on extensive software-related data, including Python codebases. Unlike rule-based tools, DeepSeek-R1 exhibits strong pattern recognition abilities for both structural and semantic changes in code.
ChatGPT-4 is a general-purpose LLM developed by OpenAI, pre-trained on diverse natural language and code corpora. Although it is not specifically designed for refactoring detection, its ability to understand complex code transformations and provide natural language explanations makes it a valuable baseline for evaluating semantic-level refactorings. 
We provide DeepSeek-R1 and ChatGPT-4 with full before-and-after file comtexts and a well-designed prompt with chain of thought (CoT) as instruction which can be found in the replication package~\cite{Replication}

\subsection{Metrics}
% Following previous work~\cite{tsantalis2020refactoringminer,opitz2019macro}, we  use four widely-used evaluation metrics including precision~\ref{con:precision},
% recall~\ref{con:recall}, Macro-F1~\ref{con:macrof1} score and Micro-F1~\ref{con:microf1} score, defined as follows:
Following previous work~\cite{tsantalis2020refactoringminer,opitz2019macro}, we  use four widely-used evaluation metrics including precision,
recall, F1 score~\cite{opitz2019macro}, defined as follows:
\begin{equation}
    \textrm{$Precision_i$} = \frac{  \textrm{$TP_i$} }{ \textrm{$TP_i$} + \textrm{$FP_i$} } \label{con:precision}
\end{equation}
\begin{equation}
    \textrm{$Recall_i$} = \frac{ \textrm{$TP_i$} }{\textrm{$TP_i$} + \textrm{$FN_i$} }
    \label{con:recall}
\end{equation}

\begin{equation}
\textrm{$F1_i$} =  \frac{2 \times \textrm{$Precision_i$} \times \textrm{$Recall_i$}}{\textrm{$Precision_i$} + \textrm{$Recall_i$}}
\label{con:f1}
\end{equation}

% \begin{equation}
% \textrm{$Macro-F1$} = \frac{1}{N} \sum_{i=1}^{N} \frac{2 \times \textrm{$Precision_i$} \times \textrm{$Recall_i$}}{\textrm{$Precision_i$} + \textrm{$Recall_i$}}
% \label{con:macrof1}
% \end{equation}

% \begin{equation}
% \textrm{$Micro-F1$} = \frac{2 \times \sum_{i=1}^{N} \textrm{$TP_i$}}{2 \times \sum_{i=1}^{N} \textrm{$TP_i$} + \sum_{i=1}^{N} \textrm{$FP_i$} + \sum_{i=1}^{N} \textrm{$FN_i$}}
% \label{con:microf1}
% \end{equation}

where \textrm{$TP_i$} is the number of valid refactoring operations mined by a tool on refactoring type $i$, \textrm{$FP_i$} is the number of invalid refactoring operations reported by a tool on refactoring type $i$, {$FN_i$} is the number of valid refactoring operations missed by a tool on refactoring type $i$, and $N$ is the total number of refactoring types.
 Precision indicates the proportion of detected refactorings that are accurate, while recall measures the extent of actual refactorings detected. The F1-score provides a balanced measure of these two aspects. 
\subsection{Dataset}

The original dataset is derived from Python-Adaptive-RefactoringMiner~\cite{dilhara2022discovering} and PyRef~\cite{atwi2021pyref}, which is available in our replication package~\cite{Replication}.
We selected all commits from these datasets that contain refactoring types supported by \appname in Table~\ref{tab:actref-rules}. 
We focus on a subset of common and structurally observable refactoring types, including class, method, and variable-level operations such as Rename, Move, Extract, and Inline. These types are well-supported by our action-based analysis framework and can be reliably detected through static structural and contextual changes.
In contrast, we exclude certain behavior-affecting or inheritance-sensitive refactorings, such as Pull Up/Push Down Method or Add/Remove/Rename Parameter. Detecting these operations in Python is challenging and harmful due to the language’s dynamic typing and flexible inheritance mechanisms~\cite{noei2025detecting}. Moreover, changes to method parameters often imply API evolution or semantic modifications, which are difficult to assess purely through static analysis and may exceed the scope of structural refactoring.
As a result, we construct a dataset of 500 commits drawn from 136 open-source Python projects. 

% We select all refactoring types supported by \appname from these two sources, resulting in a dataset comprising 500 commits from 136 open-source Python projects. 
To extend the oracle, we executed \appname, PyRef, DeepSeek-R1, and ChatGPT-4 on all 500 commits. We then manually validated the detected refactorings. During the manual detection phase, we directly classified results that matched the original dataset as true positives. Additionally, refactoring instances that were consistently detected by both \appname and PyRef were also considered true positives without further review. However, due to LLM's tendency to over-detect, we did not automatically classify their results as true positives, even when they overlapped with \appname or PyRef. Instead, for cases where there were differences, such as refactorings detected by LLM, newly identified refactorings, or missed refactorings compared to the original dataset annotations, we performed manual validation. The validation process followed a predefined set of rigorous rules (shown in Table~\ref{tab:actref-rules}) to ensure consistency. In rare instances of ambiguity, a second validator was consulted for confirmation.

\begin{figure}
    \centering
    \includegraphics[width=.65\linewidth]{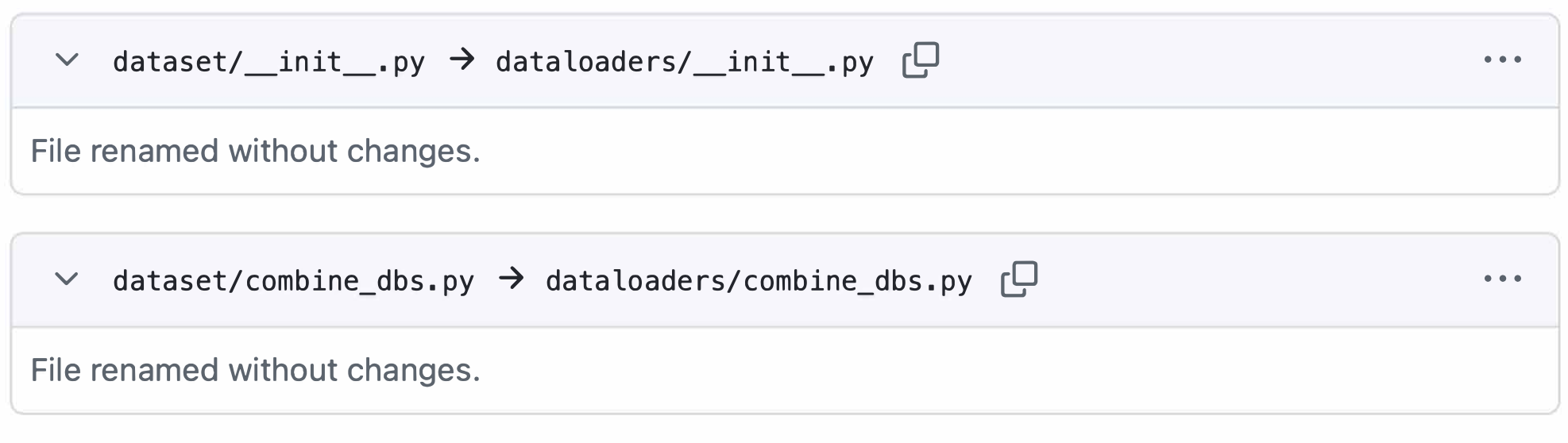}
    \caption{An Example of Move Module}
    \label{fig:model}
\end{figure}

During this process, we observed several instances of initial misclassification. 
For example, certain operations categorized as \textit{Move Method} or \textit{Move Class} were more accurately classified as \textit{Move Module}, \textit{Inline Module}, \textit{Rename Module}, or \textit{Extract Module}. Such cases were particularly common for file-level changes, where file renaming or relocation occurred without modifications to the file content (as shown in Figure~\ref{fig:model}).
These cases were reclassified based on refined definitions to better reflect their actual semantics.
% These operations were reclassified based on their actual behavior in alignment with the refined definitions. 
Additionally, new refactoring instances detected by the LLM and verified as valid through manual inspection were also incorporated. 
All discrepancies and borderline cases were resolved through a secondary review by two authors to ensure consistency.
After this thorough validation and reclassification process, we constructed a high-quality dataset containing 1,914 refactoring instances.

%% file: chapter/5-RQ.tex
\section{Results}
\label{sec:evaluation}

To evaluate the performance of \appname, we answer the following research question :

\begin{itemize}[leftmargin=*]
    \item RQ1: How does \appname perform compared to current tools for refactoring detection?
    
 This question investigates \appname's detection accuracy, comparing it to existing refactoring detection tools in terms of precision, recall, and F1 scores.
    
    \item RQ2: How does \appname's runtime compare to the state-of-the-art tool?
    
This question examines the runtime efficiency of \appname to understand whether its approach can achieve competitive performance in a practical setting.

\end{itemize}

   Each RQ is explored through carefully designed experiments that are aimed at evaluating various aspects of \appname. For each RQ, we detail the experimental setup, the metrics used for evaluation, and the results obtained.

\subsection{RQ1: How does \appname perform compared to current tools for refactoring detection?}
This research question aims to evaluate the effectiveness of \appname in detecting Python refactorings, comparing its accuracy and comprehensiveness against established refactoring detection tools. For this study, we selected rule-based tools and LLMs for comparison: \textbf{PyRef}~\cite{atwi2021pyref}, \textbf{PyRef} with \textbf{MLRefScanner}~\cite{noei2025detecting}, \textbf{DeepSeek-R1}~\cite{DeepSeek-R1}, and \textbf{ChatGPT-4}~\cite{ChatGPT-4}. 
% This prompt was designed to provide CodeLlama with both a knowledge base of refactoring types and the diff for each commit, which aids the model in identifying refactorings within code changes. However, to prevent hallucinations from very long input text, we used the commit diffs rather than the full source files, which may have resulted in some loss of structural context.
The results of different tools in terms of precision, recall, and F1-score are shown in table~\ref{tab:PRresult}.

% \begin{figure}[b]
%     \centering
%     \includegraphics[width=0.98\linewidth]{figure/F1.pdf}
%     \caption{F1 Score for \appname, PyRef and CodeLlama}
%     \label{fig:F1}
% \end{figure}

%     \begin{table}
    
% \setlength\tabcolsep{3pt}
%         \centering
%         \caption{Precision and Recall of \appname and PyRef Per Refactoring Type}
%         \label{tab:PRresult1}
%         \renewcommand{\arraystretch}{1.2}
%        \resizebox{0.95\linewidth}{!}{ 
%     \begin{tabular}{ l|c|cc|cc } 
%             \toprule
%            & & \multicolumn{2}{c}{\appname} & \multicolumn{2}{c}{PyRef}\\
%             \cmidrule(lr){3-4} \cmidrule(l){5-6}
%             Refactoring Type&Num & Precision & Recall&Precision&Recall   \\\midrule
            
%             Move Method &276& \blackwhitebar{0.786}  &\blackwhitebar{0.841} & \blackwhitebar{0.346} & \blackwhitebar{0.378}  \\
            
%             Extract Method &167& \blackwhitebar{0.558}  &\blackwhitebar{0.838} &\blackwhitebar{0.937} & \blackwhitebar{0.347} \\

%             Inline Method &38& \blackwhitebar{0.337}  &\blackwhitebar{0.789} &\blackwhitebar{0.667} & \blackwhitebar{0.244}  \\
            
%             Rename Method &432& \blackwhitebar{0.904}  &\blackwhitebar{0.977}  &\blackwhitebar{0.954} & \blackwhitebar{0.681}   \\            

%             Rename Class &207& \blackwhitebar{0.966}  &\blackwhitebar{0.952}  &\blackwhitebar{0.941} & \blackwhitebar{0.740} \\   \midrule

%                         Total &1,111& \blackwhitebar{0.783}  &\blackwhitebar{0.912} &\blackwhitebar{0.727} & \blackwhitebar{0.552} \\
%             \bottomrule
%         \end{tabular}}
%     \end{table}

    \begin{table}
    
\setlength\tabcolsep{3pt}
        \centering
        \belowrulesep=0pt
\aboverulesep=0pt
        \caption{Precision and Recall of \appname and State-of-the-Art Approaches Per Refactoring Type}
        \label{tab:PRresult}
        \renewcommand{\arraystretch}{1.2}
       \resizebox{\linewidth}{!}{ 
    \begin{tabular}{ l|c|ccc|ccc|ccc|ccc|ccc } 
            \toprule
           & & \multicolumn{3}{c|}{\appname} & \multicolumn{3}{c|}{PyRef}& \multicolumn{3}{c|}{PR+MS} & \multicolumn{3}{c|}{DeepSeek-R1} & \multicolumn{3}{c}{ChatGPT-4}\\
            \Xcline{3-5}{0.4pt} \Xcline{6-8}{0.4pt}\Xcline{9-11}{0.4pt}\Xcline{12-14}{0.4pt} \Xcline{15-17}{0.4pt}
            Refactoring Type&Num & P & R  &F1 &P  &R  &F1 & P & R& F1  &P & R& F1  &P & R &F1 \\\midrule
            
            Move Method &276  &.79  &.84 &.81  &.39 &.38 &.38  &.38  &.35 &.37  &.43  &.37  &.39  &.42  &.01  &.16\\
            
            Extract Method &167  &.56  &.84  &.67  &.94  &.35  &.51  &.93  &.30  &.45   &.58  &.77  &.67  &.46  &.72 &.57 \\

            Inline Method &38  &.34  &.79  &.47  &.67  &.24  &.36  &.67  &.24  &.36  &.28  &.57  &.38  &.08  &.05  &.06  \\
            
            Rename Method &432  &.90  &.98  &.94  &.95  &.68  &.80  &.94  &.54  &.69  &.96  &.71  &.82  &.91  &.56  &.70   \\       

            Rename Class &207  &.97  &.95  &.96  &.94  &.74  &.83  &.94  &.70  &.80  &.87  &.79  &.83   &.78  &.65  &.71 \\ \midrule
             Total (RQ1.1) &1,111& .78 &.91  &.84  &.76 &.55   &.64   &.72  &.48  &.58  &--  &--  &--  &--  &--   &--  \\ \midrule

            Move Class &34  &.94  &.91  &.93  &--   &--   &--   &--     &--   &--   &.35  &.43  &.39  &.18  &.10  &.13\\
            
            Extract Class &79  &.77  &.51  &.61  &--   &--   &--   &--  &--   &--  &.46  &.81  &.59  &.39  &.70 &.50\\

            Inline Class &1  &1.00  &1.00   & 1.00  &--   &--   &--   &--   &--   &-- &.03  &1.00  &.07  &.00  &.00  &.00\\

            Move Module &258  &1.00  &1.00  &1.00  &--   &--   &--   &--    &--   &--&.81  &.93  &.86  &.96  &.59  &.73\\
            
            Extract Module &25  &.65  &.80  &.71  &--   &--   &--   &--   &--   &--  &.12  &.47   &.19   &.19  &.23  &.20\\

            Inline Module &11  &.69   &.82   &.75  &--   &--   &--   &--   &--   &-- &.07  &.20  &.10  &.17  &.09  &.12\\
            
            Rename Module &27  &1.00  &1.00  &1.00  &--   &--   &--   &--   &--   &-- &.46  &.95  &.62  &.36  &.67  &.47\\  

            Extract Variable &88  &.70  &.86  &.78  &--   &--   &--   &--   &--   &--  &.25  &.08  &.12  &.26  &.11   &.16\\

            Inline Variable &35  &.37  &.94  &.53  &--   &--   &--   &--   &--   &-- &.06  &.03  &.04   &.06  &.03  &.04 \\
            
            Rename Variable &243& .86  &1.00  &.92 &--   &--   &--   &--    &--   &--&.44  &.61  &.51  &.46  &.34  &.39\\ \midrule

                        Total (RQ1.2)&1,921  &.80  &.92  &.85  &--   &--   &--  &--  &--   &-- &.58  &.63  &.60  &.57  &.43  &.49\\

            \bottomrule
        \end{tabular}}
    \end{table}

\vspace{1mm}
\subsubsection{RQ1-1: How does \appname perform compared to PyRef and PyRef with MLRefScanner?}

% Both \appname and PyRef are designed to capture refactorings in Python based on rules, though they differ in approach and scope. While \appname focuses on action-based analysis, PyRef builds upon RefactoringMiner2's rule-based method adapted to Python's syntax and semantics.
To assess the relative effectiveness of \appname , PyRef and PyRef with MLRefScanner (PR+MS), we only analyze their performance on the five refactoring types both tools support: \textit{Move Method}, \textit{Extract Method}, \textit{Inline Method}, \textit{Rename Method}, and \textit{Rename Class}. 

\appname consistently outperforms PyRef across five refactoring types, as reflected in its superior F1 scores of 0.84 compared to PyRef's 0.64. These higher scores indicate that \appname provides a more balanced and accurate detection capability across a broader range of refactorings. Both tools demonstrate high precision, with \appname surpassing PyRef in recall (0.91 vs. 0.55), capturing a significantly higher number of true positives in these categories. PyRef's relatively low recall suggests it may miss a substantial number of applicable refactorings, likely due to its stricter matching criteria and limited adaptability to varied code structures, which we have discussed in Chapter~\ref{sec:preliminary}. However, \appname's precision in certain categories is lower than baseline tools, such as \textit{Extract Method} and \textit{Inline Method}. 
This discrepancy can be attributed to the limitations of GumTree's SimpleMatcher algorithm. SimpleMatcher, which \appname relies on, tends to misidentify nodes with similar names, leading to false positives. 
For example, when a substantial portion of a function's body is extracted into a new method, or an old method is inlined to another method, GumTree may mistakenly interpret this \textit{Extract Method} or \textit{Inline Method} as a \textit{Rename Method} operation due to the similarity in node names and structure. 
In these cases, \appname may incorrectly report a \textit{Rename Method}, overlooking differences in parameters or calling context and focusing only on the changes in name and abstract syntax tree similarity. To mitigate this issue, we incorporated a check on the context of the action's subject, which we introduce in Chapter~\ref{sec:discussion} to reduce such false positives. Although this adjustment has helped, it has not entirely eliminated the problem. 
However, it remains valuable of \appname's detection on \textit{Inline Method} and \textit{Extract Method}, because it might be easier for developers to verify a lot of refactorings rather than identify missed refactorings.
Future improvements could focus on further enhancing context-aware analysis to improve precision in complex refactorings. Additionally, PyRef detected 26 instances that should be classified as \textit{Extract Class}, \textit{Move Class}, or other module-level refactoring, but it incorrectly identified them as \textit{Move Method}. This misclassification occurred because PyRef does not support the detection of these refactoring types. Due to this limitation, we have decided to omit these instances from PyRef's result.

In summary, \appname stands as the preferred choice for scenarios demanding thorough refactoring detection and a balance between precision and recall. Its ability to adapt to varied refactoring patterns makes it particularly suitable for large-scale, diverse datasets, providing actionable insights that surpass PyRef's narrower focus on precision.

In addition to comparing \appname with PyRef, we also evaluated the combination of PyRef and MLRefScanner (denoted as MS+PR). MLRefScanner is a classifier trained to identify whether a commit contains refactoring, and when combined with PyRef, it aims to improve detection by filtering or re-ranking results. 
However, our results show that MS+PR underperforms PyRef in precision, recall and F1 scores, highlighting that the classifier does not significantly enhance PyRef’s structural matching capabilities.
MS+PR consistently demonstrates lower recall across all five refactoring types, particularly in \textit{Extract Method} and \textit{Inline Method}, where the recall dropped to 0.30 and 0.24. These findings suggest that MLRefScanner, which is optimized for commit-level classification, is not well-suited for identifying fine-grained method-level refactorings. It may introduce noise or overly conservative filtering, resulting in missed detections. The lack of contextual understanding at the statement level limits the effectiveness of this hybrid approach. Therefore, while MS+PR may be useful for commit-level refactoring detection, it offers little advantage in detecting specific refactoring operations compared to using PyRef alone.
These observations further reinforce the advantage of \appname’s action-based strategy. Unlike MS+PR, which relies on high-quality commit messages, \appname analyzes concrete AST-level changes action, yielding a better balance between precision and recall, particularly in complex or dynamic codebases.

\subsubsection{RQ1-2: How does \appname perform compared to LLMs?}
To evaluate whether LLMs can effectively detect Python refactorings, we compared \appname against two strong LLM-based baselines: DeepSeek-R1, an open-source code model fine-tuned for reasoning, and ChatGPT-4, a state-of-the-art commercial LLM. The results, summarized in Table~\ref{tab:PRresult}, demonstrate a significant performance gap between \appname and both LLMs across all 15 refactoring types. 
It is worth noting that, due to the input length constraints of LLMs, a small number of commits could not be processed by LLMs: 21 commits exceeded the token limit for DeepSeek-R1, and 5 for ChatGPT-4. We excluded the out-of-limit-token commits from the evaluation of referenced LLM.
\appname achieves an F1 score of 0.85, outperforming DeepSeek-R1 (F1: 0.60) and ChatGPT-4 (F1: 0.49) by large margins. In terms of recall, \appname achieves 0.92 compared to 0.63 and 0.43 for DeepSeek-R1 and ChatGPT-4. This shows that \appname is far more capable of recovering actual refactoring instances. Moreover, even in terms of precision, where LLMs are often assumed to excel due to their semantic understanding, \appname still outperforms both (P: 0.80 vs. 0.58 and 0.57). These findings suggest that current LLMs tend to miss many refactorings (i.e., low recall) and sometimes hallucinate patterns that do not correspond to actual structural changes (moderate precision).

We also observe that \appname’s precision or recall is relatively lower in a few categories (e.g., Extract Method and Extract Class). 
Extract Method's precision is lower than DeepSeek-R1 (0.56 vs. 0.58),
 recall drops to 0.51 in Extract Class, indicating that \appname may miss extractions with significant transformations or multi-step restructuring. Nevertheless, the F1 remains reasonable (0.61), demonstrating that most detected cases are still correct.

Despite such challenges, \appname consistently outperforms both LLMs on every single refactoring type in terms of F1 score. LLMs like DeepSeek-R1 and ChatGPT-4 exhibit inconsistent behavior: they may achieve relatively high recall in a few categories (e.g., ChatGPT-4 on Extract Class), but often at the cost of extremely low precision. This leads to significantly lower F1 scores, suggesting that LLMs still struggle to distinguish refactoring patterns from general edits in code.

\begin{center}
\begin{tcolorbox}[colback=gray!10,%gray background
                  colframe=black,% black frame colour
                  width=13cm,% Use 8cm total width,
                  arc=1mm, auto outer arc,
                  boxrule=0.5pt,
                 ]
  \textbf{Answer to RQ1}:
\appname consistently outperforms both rule-based tools and LLMs in detecting Python refactorings across a range of types. \appname's higherF1 suggest that it provides more balanced and accurate detection.
\end{tcolorbox}
\end{center}

% \appname demonstrates superior performance over both PyRef and CodeLlama in detecting Python refactorings, achieving a more balanced precision and recall across various refactoring types, making it a highly effective tool for real-world applications.
% \end{tcolorbox}

% \end{center}

\begin{figure}[htbp]

		\centering
		\includegraphics[width=.6\textwidth]{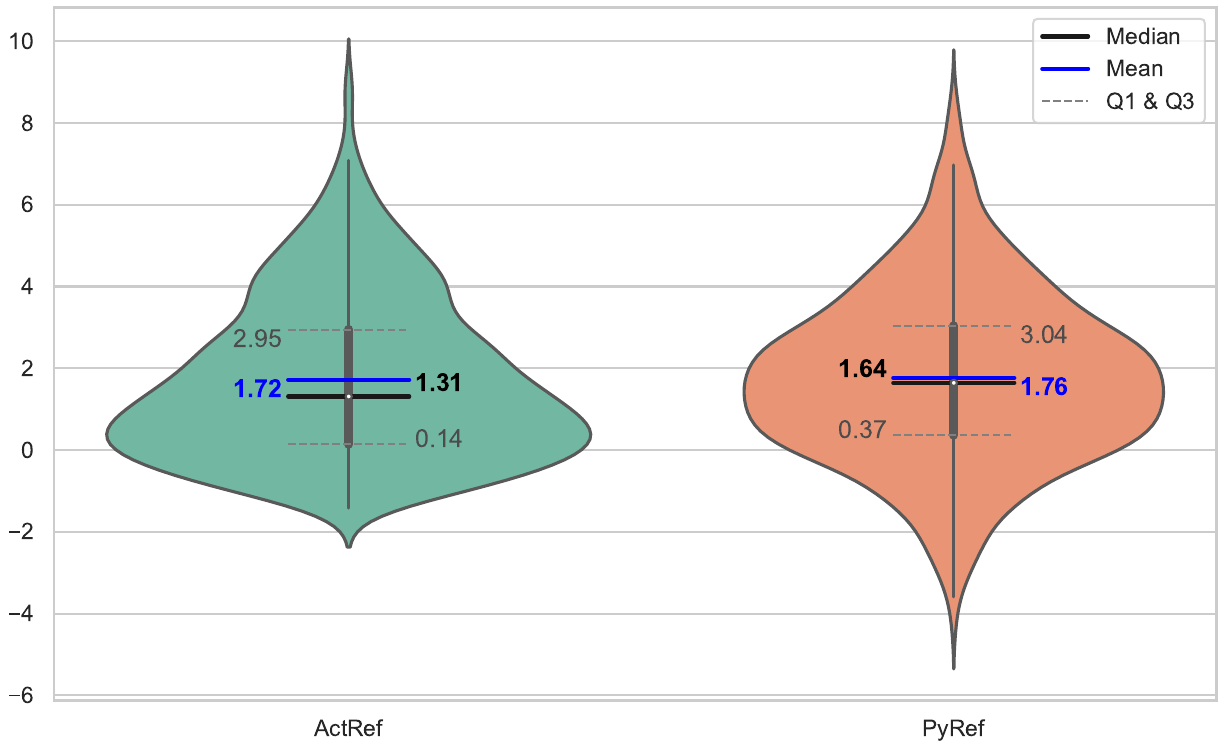}
        \caption{Runtime Comparison of \appname and PyRef}
        \label{fig:runtime}

\end{figure}
\subsection{RQ2: How does \appnamebold's runtime compare to state-of-the-art tool?}
In this research question, we aim to evaluate the runtime performance of \appname in comparison to PyRef.
We conducted an experiment using the same dataset in RQ1, which contains commits from multiple open source Python projects, covering different domains and sizes. We executed separately each tool on the same machine with the following specifications: Intel Xeon Platinum 8358P CPU 2.60GHz, 2.0 TiB of DDR4 memory, running Ubuntu 20.04.6 LTS and Python 3.7.
In order to avoid the impact of network fluctuations on the experiment, we downloaded all the Python codes involved in the commit to the local computer in advance, so we bypassed the git parsing module when executing PyRef. For each commit in the dataset, both \appname and PyRef were executed and the time taken to complete the analysis was measured. This approach provides a precise measurement of how each tool performs on individual commits, avoiding aggregate measurements that could obscure commit-level performance variations. Each commit was analyzed three times to reduce the impact of random fluctuations in runtime caused by external factors such as system load. The average runtime across these trials was then calculated for each tool, ensuring that the reported results reflect consistent performance.

% \begin{figure}[h]
%     \centering
%     \includegraphics[width=0.98\linewidth]{figure/runtime.pdf}
%     \caption{Runtime Comparison of \appname and PyRef (Log Scale)}
%     \label{fig:runtime}
% \end{figure}

We use violin plots to illustrate the runtime distribution of \appname and PyRef. To intuitively show the runtime performance of \appname and PyRef on all test cases, we logarithmically transformed the runtime. As shown in Figure~\ref{fig:runtime}, although \appname supports more refactoring types (15, compared to 10 for PyRef), the runtimes of the two are very close in terms of average and median. This shows that \appname can maintain comparable performance to the existing tool while supporting more refactoring types' detection. The distribution of \appname has some large outliers at the top, which are mainly due to the fact that \appname supports more complex refactoring types, such as cross-file refactoring or more complex extract and move operations. The detection of these operations requires more computational resources, thus resulting in longer runtimes in some more complex code commits. \appname uses the SampleMatch-gumtree action calculation algorithm, which can lock the changed areas in the code more quickly and accurately to avoid redundant operations. This is particularly critical in complex refactoring scenarios (such as extraction, movement, and cross-file operations), so even if more refactoring types are supported, the running time of \appname is still similar as PyRef.

\begin{center}
\begin{tcolorbox}[colback=gray!10,%gray background
                  colframe=black,% black frame colour
                  width=13cm,% Use 8cm total width,
                  arc=1mm, auto outer arc,
                  boxrule=0.5pt,
                 ]
  \textbf{Answer to RQ2}:
\appname matches PyRef in runtime performance while supporting a broader range of refactoring types and finer detection granularity. This balance suggests that \appname can deliver comprehensive refactoring insights without added runtime cost, enhancing its practicality for detailed, real-world code analysis.
\end{tcolorbox}
\end{center}

%% file: chapter/6-discussion.tex
\section{Discussion}
\label{sec:discussion}
In this section we provide a comprehensive discussion of the research, examining both the implications and potential threats. The implications section highlights the broader impact of \appname on Python code refactoring detection, while the threats section addresses possible limitations, ensuring a balanced view of the results and their reliability.
\subsection{Implications}
The implications of our work are summarized as follows:
\begin{itemize}[leftmargin=*]
\item The introduction of an action-based analysis approach to refactoring behaviours enables the observation and evaluation of these behaviours from a new perspective, while also revealing the complexity and diversity of refactoring operations. This novel perspective helps to address the current knowledge gap in refactoring analysis within the academic community, offering a theoretical foundation for future research.
\item The staged analysis pipeline from module-level to AST-level actions demonstrates a robust and scalable strategy for detecting a wide range of refactorings, from coarse-grained refactoring to fine-grained refactoring. This layered design not only improves detection accuracy but also enhances extensibility, making it suitable for analyzing large-scale, real-world software projects. 
\item  \appname is inherently extensible, offering a unified representation for detecting diverse refactoring types. As new refactoring patterns emerge, they can be accommodated by augmenting the rule set without redesigning the entire system. This design paves the way for building more intelligent, adaptable refactoring detectors across programming languages and ecosystems. Extending the approach taken by \appname to adapt to other programming languages or frameworks is our ongoing work.
\end{itemize}
% (1) The introduction of an action-based analysis approach to refactoring behaviours enables the observation and evaluation of these behaviours from a new perspective, while also revealing the complexity and diversity of refactoring operations. 
% This novel perspective helps to address the current knowledge gap in refactoring analysis within the academic community, while also providing a theoretical basis for future research in this field. 
% (2) The staged analysis pipeline from module-level to AST-level actions demonstrates a robust and scalable strategy for detecting a wide range of refactorings, from coarse-grained refactoring to fine-grained refactoring. This layered design not only improves detection accuracy but also enhances extensibility, making it suitable for analyzing large-scale, real-world software projects. 
% (3) \appname is inherently extensible, offering a unified representation for detecting diverse refactoring types. As new refactoring patterns emerge, they can be accommodated by augmenting the rule set without redesigning the entire system. This design paves the way for building more intelligent, adaptable refactoring detectors across programming languages and ecosystems. Extending the approach taken by \appname to adapt to other programming languages or frameworks is our ongoing work.

\subsection{Internal Threats}
The internal threat comes from the choice of the action calculation algorithm. Although GumTree has shown reliable performance in previous studies and our tests, it is not flawless and may sometimes misclassify changes, resulting in false positives. The detected action can vary significantly for the same code snippets based on the selected threshold~\cite{aniche2016satt,siefert2021evaluation}. Although we tested alternative strategies of GumTree (ClassicMatcher and GreedyMatcher) and thresholds for calculating actions, the experimental results indicated that SimpleMatcher with a self-adaption threshold offered the most balanced trade-off between accuracy and computational efficiency and comprehensive results for our specific use case. We also implement several filters to reduce the risk of such misclassifications, yet some level of error remains inevitable. Future improvements may involve integrating other matching algorithms to further enhance the precision of action detection. Still, the current use of SimpleMatcher paired with our context-aware post-processing mechanism provides a robust foundation for this study's findings.

GumTree tends to generate the shortest edit script by maximizing subtree matches, which may cause false positives in Rename Method and false negatives in Extract Method when large code fragments are extracted. To mitigate this, we apply post-processing rules that verify whether the original node has been fully or partially deleted, and whether similar fragments are added elsewhere, before classifying a change as an update or a move. Additionally, we enhanced GumTree to consider semantic signatures when pairing top-level declarations, which mitigates the common false positives in Inline Method and Extract Method cases that arise from similar subtree structures.

Additionally, we initially considered using Python-Adaptive-Refactoring Miner~\cite{dilhara2022discovering} and were able to successfully execute its replication package with assistance from the original authors, we ultimately excluded it from our evaluation due to significant reliability issues. Specifically, the tool failed to detect any refactoring operations, even when applied to the original dataset provided by the authors. Furthermore, this limitation is not unique to our experience, similar issues have been reported by other users on the tool’s official replication package repository\footnote{https://github.com/mlcodepatterns/PythonTypeInformation/issues/2}, suggesting a broader reproducibility problem. 

\subsection{External Threats}

A key external threat to the validity of our results lies in the scope and representativeness of the dataset. Our dataset is drawn from previous work~\cite{atwi2021pyref, dilhara2022discovering}, which identified refactorings in real-world development scenarios. These datasets were curated using state-of-the-art detection tools, ensuring that the commits represent authentic refactoring practices in open-source projects. However, as the initial datasets only include commits with refactorings detected by prior tools, they are inherently biased toward simpler refactoring cases. This bias stems from the limitations of the detection tools themselves, which often struggle to identify more complex or nuanced refactoring patterns, resulting in an overrepresentation of straightforward scenarios.

To address this limitation and enrich the dataset, we performed additional detection on the original commits, uncovering and incorporating new refactoring instances that were not initially identified. While this process expanded the dataset and increased its diversity, the overall composition might still lean towards simpler refactoring cases, as the starting point was shaped by the detection capabilities of the tools used in earlier works. This tendency toward simpler cases highlights a broader challenge in refactoring research: mainstream detection tools frequently perform well on well-defined refactorings but face significant difficulties in detecting more intricate or multi-step transformations. Consequently, our evaluation, while reflective of the characteristics of the dataset, might not fully capture the performance of the tools in complex real-world scenarios.
Future work should focus on extending the dataset to better represent the full spectrum of refactoring practices in software development, including those involving intricate transformations and interactions across multiple files or modules. Such efforts would provide deeper insights into the challenges faced by detection tools and further enhance the robustness of their evaluation.
% five large, popular, and actively maintained Python-based machine learning projects and data process projects. While these projects provide a wealth of refactoring examples and diversity in development practices, their prominence and size may introduce a bias toward specific types of refactorings that are more common in large-scale or highly modular projects. This might limit the generalizability of our findings when applied to smaller or less active projects, where the frequency, type, and context of refactorings could differ significantly. To address this, we selected projects that are representative of widely used open-source software, believing that these projects reflect a broad range of refactoring scenarios. Nevertheless, further validation across different project types is necessary to fully understand the tool's performance across diverse development environments.

Another external threat is the nature of using action-based detection for identifying refactorings, especially when compared to established tools like PyRef. While action-based detection allows for a more flexible and human-centric analysis of code transformations, it may struggle with certain refactorings that involve more nuanced semantic changes not captured in the action set. This could result in some refactorings going undetected. While our approach aims to overcome some limitations of previous tools by focusing on both structural and semantic aspects of code changes, future work should include the evaluation of our tool on more diverse datasets and across different programming languages to assess its broader applicability. 

%% file: chapter/7-relatedwork.tex
\section{Related Work}
\label{sec:relatedwork}
Weißgerber and Diehl~\cite{weissgerber2006identifying} introduced a pioneering technique for detecting local-scope and class-level refactorings in the commit histories of CVS repositories. Their approach begins by identifying potential refactoring candidates through pairs of code elements (e.g., classes, methods, fields) that exhibit signature differences. To analyze these candidates further, they employed the clone detection tool CCFinder~\cite{kamiya2002ccfinder}, configured to tolerate variations such as whitespace and comment changes, as well as consistent renaming of variables, methods, and member references. To evaluate their method, they manually inspected commit log messages from two open-source projects to identify documented refactorings and calculate recall. Additionally, they used random sampling to estimate the precision of their technique.

Ref-Finder~\cite{kim2010ref} utilizes a logic programming approach to detect a wide range of refactorings, offering the most comprehensive coverage among tools based on Fowler's catalog, supporting 63 refactoring types. It defines each refactoring as a set of logic query templates that capture structural constraints before and after the refactoring. This approach excels at identifying both atomic (e.g., Rename Method) and composite (e.g., Extract Superclass) refactorings, making it highly versatile for Java projects. However, its reliance on logic-based templates and a formalized structure heavily ties it to Java's statically typed environment, which limits its application to dynamic languages like Python, where structure may be less strictly defined. Moreover, Ref-Finder's approach can become computationally expensive and less practical for real-world commits that mix refactorings with bug fixes or feature additions. Its dependency on the correctness of structural facts means that overlapping or noisy changes are harder to filter out, making it less suitable for detecting refactorings in mixed or complex commits.

JDevAn~\cite{Xing_Stroulia_2006, xing2008jdevan} proposed a method for detecting refactorings by analyzing fine-grained changes in Abstract Syntax Trees across commits. Their approach is based on the design-level changes reported by UMLDIFF~\cite{xing2005umldiff}, which is a domain-specific structural differencing algorithm that analyzes two class models, reverse-engineered from consecutive versions of an object-oriented system, to identify structural changes in packages, classes, interfaces, fields, and methods. The authors evaluated their technique by validating detected refactorings in several open-source Java projects, demonstrating its effectiveness in capturing a broad range of documented refactorings.

RefactoringMiner 2.0~\cite{tsantalis2020refactoringminer} is a state-of-the-art tool for refactoring detection focused primarily on Java. It is renowned for its precision and recall, achieving a high accuracy without relying on code similarity thresholds, unlike many of its predecessors. The tool specializes in detecting both high-level and low-level refactorings, including Extract Method and Rename Method, as well as submethod-level refactorings such as Extract Variable. This makes it particularly effective for Java projects. However, RefactoringMiner's reliance on Java's static type system and specific statement-matching techniques makes it language-specific, limiting its direct applicability to dynamic languages like Python. The approach is tied closely to object-oriented constructs typical of statically typed languages, which makes it harder to generalize to dynamic languages that might not have strict type declarations or consistent structural patterns.

Refdiff 2.0~\cite{silva2020refdiff} is a multi-language refactoring detection tool designed to overcome the limitation of being restricted to a single language, such as Java, by providing support for Java, JavaScript, and C. 
RefDiff abstracts code structure into a Code Structure Tree (CST), which enables it to identify high-level refactorings, including renames, moves, and extracts. However, while it can generalize across multiple languages, RefDiff is still reliant on a token-based similarity measure that may miss more granular or subtle changes, such as method body refactorings (like Extract Variable) that occur frequently in dynamic languages like Python. RefDiff focuses on detecting low-level refactorings and handling more complex code changes common in languages like Python, makes it less suited for cases where a detailed, intra-method refactoring analysis is needed, or when there are overlapping changes that mix refactoring with other types of modifications

Python-Adaptive-RefactoringMiner~\cite{dilhara2022discovering}, adapted from RefactoringMiner, focuses on detecting refactorings in Python code by transform Python AST to Java AST and building on the strengths of RefactoringMiner's statement-matching techniques. It retains high precision for detecting certain refactorings, such as method extraction and renaming, through statement matching algorithms. However, like RefactoringMiner, this tool struggles with Python's dynamic nature. Python lacks the strict typing and structural rigidity found in Java, making it harder for tools like this to precisely capture refactorings when code undergoes significant structural changes or dynamic behavior alterations. Additionally, Python-Adaptive-RefactoringMiner may be challenged by real-world Python commits, which often blend functional changes with refactorings. Its statement-matching algorithm, while effective for straightforward refactorings, might misinterpret or miss complex refactorings embedded in larger modifications.

PyRef~\cite{atwi2021pyref}, based on the RefactoringMiner framework, focuses on detecting a limited set of refactorings. While these refactorings are essential, PyRef's limited scope restricts its ability to handle the wide variety of refactorings often seen in real-world Python projects, which may involve more intricate or composite changes. PyRef's reliance on AST-based matching methods poses challenges when handling Python's dynamic features, such as duck typing, flexible function signatures, and frequent structural changes. This limitation becomes particularly pronounced in commits that blend refactoring with other functional changes, leading to a reduced ability to detect complex refactorings across multiple files or in larger codebases.

MLRefScanner~\cite{noei2025detecting} is a prototype tool that applies ML techniques to detect refactoring commits in ML Python projects. MLRefScanner extracts textual, process, and code features to train classifiers to detect commits. They conducted extensive experiments on 199 ML Python projects and verified the performance of MLRefScanner in various testing scenarios, and achieves a precision of 94\%, a recall of 82\%, and an AUC of 89\%. However, MLRefScanner relies on high-quality commit messages and is unable to detect specific types of refactorings.

Existing refactoring detection tools like RefactoringMiner, RefDiff, and Ref-Finder offer effective mechanisms for identifying code changes, but they face limitations in Python-specific code changes. PyRef provides high precision but support only a limited set of refactorings and struggle with Python's flexible code structure and dynamic nature. RefDiff, though multi-language, relies on code similarity, making it less effective for complex or mixed changes in real-world commits. These limitations highlight the need for a more flexible, action-based approach that better captures nuanced refactorings, especially in dynamic languages like Python, where existing methods fall short.

%% file: chapter/8-conclusion.tex
\section{Conclusion}
\label{sec:conclusion}
This paper introduces \appname, a novel refactoring mining algorithm based on code change actions, aiming to enhance the understanding of Python code refactoring. In contrast to existing methodologies, \appname reveals the diversity and complexity of refactoring operations by analyzing specific actions within the code change process, thus offering a novel analytical perspective to software engineering. Experimental results indicate that \appname outperforms current advanced refactoring detection tools in both effectiveness and efficiency, excelling in identifying complex code change patterns and distinguishing among various refactoring types. \appname also demonstrates superior accuracy and detail in refactoring identification compared to large language models, highlighting the value of structured action-based methods in consistently capturing code changes with precision. 

In conclusion, \appname introduces a new perspective on understanding code refactoring, while opening up pathways for future research. Our ongoing work is to explore \appname's efficacy in multi-language settings and advance the automation of action analysis through deep learning to address increasingly complex code evolution needs.